\documentclass[aps,prm,preprint,showkeys,preprintnumbers,amsmath,amssymb]{revtex4-1}
\usepackage{hyperref}
\usepackage{amsmath,amssymb,multirow,rotating,color,graphicx,blindtext,wrapfig,lipsum}
\usepackage{float}
\newcommand*{\impuls}[3]{p^{#1}_{#2#3}(\vec{k})}
\begin{document}
\title{Nonlinear optical response of ferroelectric oxides: 
first-principles calculations within the time-domain and the frequency-domain}

\author{Christof Dues$^{1}$}
\author{Marius J. M\"uller$^2$}
\author{Sangam Chatterjee$^2$}
\author{Claudio Attaccalite$^{3}$} 
\author{Simone Sanna$^4$} \email{simone.sanna@theo.physik.uni-giessen.de}
\scriptsize
\affiliation{$^1$ Institut f\"{u}r Theoretische Physik and Center for Materials Research (LaMa), Justus-Liebig-Universit\"{a}t Gie\ss{}en, Heinrich-Buff-Ring 16, D-35392 Gie\ss{}en, Germany}
\affiliation{$^2$ Institute of Experimental Physics I and Center for Materials Research (LaMa), Justus-Liebig-Universit\"{a}t Gie\ss{}en, Heinrich-Buff-Ring 16, D-35392 Gie\ss{}en, Germany}
\affiliation{$^3$ CNRS/Aix-Marseille Universit\'{e}, Centre Interdisciplinaire de Nanoscience de Marseille UMR 7325 Campus de Luminy, 13288 Marseille Cedex 9, France}

\date{\today}

\normalsize

\begin{abstract}
The second and third order nonlinear susceptibilities of the ferroelectric oxides LiNbO$_3$, LiTaO$_3$, and KNbO$_3$ are calculated from first principles. 
Two distinct methodologies are compared, one approach is based on a perturbative approach within the frequency-domain, another on the time-evolution of the electric polarization. 
The frequency dependence of the second harmonic coefficients of the ferroelectric phase of LiNbO$_3$ calculated within the two approaches is in excellent agreement. 
This is further validated by experimental data for LiNbO$_3$ and LiTaO$_3$, measured for an incident range of photon energies between 0.78\,eV and 1.6\,eV.
The real-time based approach is furthermore employed to estimate the third order nonlinear  susceptibilities of all investigated ferroelectric oxides. 
We further show that the quasiparticle effects, considered by means of a scissors-shift in combination with the the computationally efficient independent particle approximation, result in a shift all spectral features towards higher energies and decrease the magnitude of the optical nonlinearities. 
The energy of the main resonances in the hyperpolarizabilities suggests that the spectra can be understood by multi-photon adsorption within the fundamental bandgap for all investigated materials.
\end{abstract}

\maketitle

\section{Introduction} \label{sec:introduction}
Nonlinear optical phenomena are at the basis of many technological applications \cite{boyd2020nonlinear,Ency,shen1984principles,butcher1990elements}. 
They range from,  e.g., well established frequency converters \cite{PhysRev.178.2036} to exciting novel approaches for white light generation \cite{alfy,jacs6b10738,scif6138,rev130,Kathy,Dornsiepen2019,Kev}, which is currently understood as a combination of several nonlinear optical effects of different order \cite{alfano1970emission,Book2012}. 
The demand for a theoretical description of nonlinear effects and the accurate calculation of the related susceptibilities has grown parallel to the employment of nonlinear optical materials in different devices.

Unfortunately, the \textit{ab initio} modelling of the nonlinear optical response of a medium is one of the most challenging tasks in theoretical physics. 
While the linear-response optical properties are successfully calculated from first principles within Green's function theory \cite{oni2}, this is not the case for nonlinear optical susceptibilities.
The state-of-the-art approach for the computation of the linear optical response combines band structures calculated in many-body perturbation theory \cite{Aryasetiawan_1998} (e.g., by $G_0W_0$) to include quasiparticle effects, with the solution of the Bethe-Salpeter equation to account for the electron-hole attraction \cite{Strinati88}. 

Within this approach, it is difficult to elaborate expressions for the nonlinear optical susceptibilities which include the many-body effects \cite{zach}.
Accurate second-order susceptibilities can be calculated, e.g., on the basis of the electronic wavefunctions from a combination of density functional and a perturbation theory \cite{botti,Leitsmann2005}.
However, the complexity of the expressions derived within perturbation theory quickly grows with the perturbation order \cite{Arthy17}, and makes nonlinearities of higher orders \textit{de facto} inaccessible \cite{Jsipe}.
The applications are generally limited to smaller systems such as periodic 
crystals \cite{zach}. Up to our knowledge, a single attempt to solve the Bethe-Salpeter 
equation for second harmonic generation can be found in the literature \cite{Arthy17}.

The calculation of the nonlinear optical susceptibilities can be also performed in the time domain from the dynamical polarization \cite{Yab4484,Takim,Castro80,2960628}. 
The response of a medium to a time-dependent electrical field may be expanded into a power series:

\begin{equation}
\label{eq:potenzreihe_P}
\begin{aligned}
P_\alpha (\omega) & = \sum_{\beta}\chi^{(1)}_{\alpha\beta}(-\omega;\omega)E_{\beta}(\omega)  \\
                  & + \sum_{\beta\gamma}\chi^{(2)}_{\alpha\beta\gamma}[-\omega=-(\omega'+\omega''); \omega', \omega'']E_{\beta}(\omega')E_{\gamma}(\omega'')  \\ 
                  & + \sum_{\beta\gamma\delta}\chi^{(3)}_{\alpha\beta\gamma\delta}[-\omega=-(\omega'+\omega''+\omega'''); \omega', \omega'',\omega''']E_{\beta}(\omega')E_{\gamma}(\omega'')E_{\delta}(\omega''') \\
                  & + \cdots .
\end{aligned}
\end{equation}

Thus, if the dynamical polarization is calculated, e.g., by numerical integration of the equations of motion in presence of a laser field, optical susceptibilities of virtually 
any order can be obtained. 
Moreover, several nonlinear phenomena such as sum- and difference-frequency generation or four-wave mixing can be calculated simultaneously, as they are described by the
same equations of motion. 
Thereby, the dynamical polarization must be calculated as a geometric Berry phase as described in the modern theory of polarization \cite{Resta1994a}, if periodic boundary conditions are applied; this is generally required for the description of crystalline solids. 
In the past, one of the authors has presented a practical implementation of  this approach \cite{Attaccalite2011}, in which the equations of motion are derived  following the scheme introduced by Souza \textit{et al.}, \cite{Souza2004} based on the generalization of the Berry phase to the dynamical polarization \cite{Resta2007,Resta1994}. 

In the present work, we calculate and the nonlinear optical susceptibilities of different ferroelectric oxides both within the frequency-domain and within the time-domain
approach and corroborate the results with cooresponding experimental data. 
In a first step, we calculate the second harmonic generation (SHG) spectrum of LiNbO$_3$ up to energy of the incoming photons of 6\,eV, based both on the momentum  matrix approach and on the time evolution of the polarization. 
The excellent agreement of the spectra calculated by the different methods validates both approaches against each other, furthermore confirming earlier theoretical and experimental results \cite{Riefer13}.
In a second step, we calculate the SHG spectrum as well as the third harmonic generation (THG) spectra of LiNbO$_3$, LiTaO$_3$ as well as KNbO$_3$, and, if possible, compare them with the experimental results and calculations from the literature. 
Furthermore, we investigate the role of quasiparticle effects  on the nonlinear optical response by making use of the computationally efficient  independent particle approximation and incorporating quasiparticle effects by a previously determined scissors-shift.
In addition, we show that the major resonances in the nonlinear spectra can be explained by multi-photon processes within the fundamental bandgap.

\section{Methodology} \label{sec:theory}

\subsection{Second harmonic generation in the frequency-domain: the momentum matrix approach}

In this work, we employ the perturbative approach explained in detail in Refs.~\cite{Leitsmann2005,Riefer13}.
The method is based on the calculation of the momentum matrix elements 

\begin{equation}
\impuls{\alpha}{n}{m} = \left\langle n\vec{k} \middle| \hat{p}^\alpha \middle| m\vec{k}\right\rangle 
\end{equation}

where $\left| m\vec{k}\right\rangle$ and $\left| n\vec{k}\right\rangle$ are two Bloch states.

Considering the transition energies $\hbar\omega_{n,m}=\varepsilon_{m\vec{k}}-\varepsilon_{n\vec{k}}$  between the states $m$ and $n$ at the reciprocal space point $\vec{k}$, and the notation (derived from the anticommutator of the momentum operators in the cartesian directions $\beta$ and $\gamma$)

\begin{equation}
\left\lbrace\impuls{\beta}{m}{l}\impuls{\gamma}{l}{n}\right\rbrace = \frac{1}{2}\left[\impuls{\beta}{m}{l}\impuls{\gamma}{l}{n}+\impuls{\gamma}{m}{l}\impuls{\beta}{l}{n} \right],
\end{equation}

the SHG susceptibility is calculated for a complex frequency $\tilde{\omega}=\omega+i\eta$, in which the small positive imaginary part $i\eta$ adiabatically switches the electromagnetic
field on. 
In our work, we set $\eta$ to 0.2\,eV for all the calculations.

The expression for the SHG susceptibility reads

\begin{equation}
\label{total_chi}
\begin{aligned}
\chi_{\alpha\beta\gamma}^{(2)}(-2\omega;\omega,\omega)= & -\frac{ie^3}{\tilde{\omega}^3\hbar^2m^3V}\sum_{\vec{k}}\sum_{nml} \frac{1}{\left[\omega_{mn}(\vec{k})-2\tilde{\omega}\right]} \times \\ 
& \left[\frac{f_{nl}(\vec{k})\impuls{\alpha}{n}{m}\left\lbrace\impuls{\beta}{m}{l}\impuls{\gamma}{l}{n} \right\rbrace}{\omega_{ln}(\vec{k})-\tilde{\omega}}\right. 
		+\left.\frac{f_{ml}(\vec{k})\impuls{\alpha}{n}{m}\left\lbrace\impuls{\gamma}{m}{l}\impuls{\beta}{l}{n} \right\rbrace}{\omega_{ml}(\vec{k})-\tilde{\omega}}\right] .
\end{aligned}
\end{equation}

In this expression, the eigenvalues $\varepsilon_{n\vec{k}}$ can be evaluated either within the independent Partice Approximation (IQA, i.e., DFT with (semi)local xc-potentials) or within the independent quasiparticle approximation (IQA, e.g., from $G_0W_0$ calculations).

Equation \ref{total_chi} can be divided into a two-band term

\begin{equation}
\label{two_bands}
\begin{aligned}
\chi_{\alpha\beta\gamma}^{(2),\mathrm{two}}(-2\omega;\omega,\omega)= & -\frac{ie^3}{\hbar^2m^3V}\sum_{\vec{k}}\sum_{nm} \left[\frac{16 f_{nm}(\vec{k})\impuls{\alpha}{n}{m}\left\lbrace \Delta_{mn}^\beta (\vec{k}) \impuls{\gamma}{m}{n} \right\rbrace}{\left[\omega_{mn}(\vec{k})\right]^4 \left[\omega_{mn}(\vec{k})-2\tilde{\omega}\right]} \right. \\ 
 & - \left. \frac{f_{nm}(\vec{k})\impuls{\alpha}{n}{m}\left\lbrace \Delta_{mn}^\beta (\vec{k}) \impuls{\gamma}{m}{n} \right\rbrace}{\left[\omega_{mn}(\vec{k})\right]^4 \left[\omega_{mn}(\vec{k})-\tilde{\omega}\right]}\right] 
\end{aligned}
\end{equation}		

and a three-band term

\begin{equation}
\label{three_bands}
\begin{aligned}
\chi_{\alpha\beta\gamma}^{(2),\mathrm{three}}(-2\omega;\omega,\omega)= & -\frac{ie^3}{\hbar^2m^3V}\sum_{\vec{k}}\sum_{\substack{nml \\n\neq m\neq l}}\frac{\impuls{\alpha}{n}{m} \left\lbrace \impuls{\beta}{m}{l}\impuls{\gamma}{l}{n} \right\rbrace  }{\omega_{ln}(\vec{k})-\omega_{ml}(\vec{k})} \times \\
& \left[  \frac{16f_{nm}(\vec{k})}{\left[\omega_{mn}(\vec{k})\right]^3\left[\omega_{mn}(\vec{k})-2\tilde{\omega}\right]} \right. + \left. \frac{f_{ml}(\vec{k})}{\left[\omega_{ml}(\vec{k})\right]^3\left[\omega_{ml}(\vec{k})-\tilde{\omega}\right]} \right. \\
& \left. + \frac{f_{ln}(\vec{k})}{\left[\omega_{ln}(\vec{k})\right]^3\left[\omega_{ln}(\vec{k})-\tilde{\omega}\right]} \right] .
\end{aligned}
\end{equation}

In equation \ref{two_bands}, the matrix elements of the intraband transitions $\Delta_{mn}^\beta(\vec{k})=p_{mm}^\beta(\vec{k})p_{nn}^\beta(\vec{k})$ are calculated as

\begin{equation}
p_{mm}^\beta(\vec{k})=\frac{m_e}{\hbar}\partial_{k_\beta}\varepsilon_n(\vec{k}) .
\end{equation}

\subsection{Hyperpolarazibilites in the time domain: Time evolution of the polarization}

For the calculation of the nonlinear optical response in time-domain from the time evolution of the polarization, we employ the procedure described in Ref.~\cite{Attaccalite2013}, which we briefly outline in the following.
The starting point are the (zero-field) Kohn-Sham equations of the form,

\begin{equation}
\label{kohnsham}
\hat{H}^{0, \rm IPA}=-\frac{\hbar^2}{2m}\sum_i{\nabla_i^2+\hat{V}_{\rm eI}+\hat{V}_{\rm H}[\rho_0]+\hat{V}_{\rm xc}[\rho_0]} 
\end{equation}

where ${V}_{\rm eI}$ is the electron-ion interaction and ${V}_{\rm H}$ and $\hat{V}_{\rm xc}$ are the Hartree and the exchange-correlation potentials, respectively. 
Improvements beyond the IPA might be introduced at this step to consider quasiparticle shifts or effects originating from the response of the effective potential to density fluctuations in a time-dependent screened Hartree-Fock manner.
The corresponding Hamiltonian is denoted in this case as $\hat{H}^{0}$ instead of $\hat{H}^{0, \rm IPA}$.

The laser excitation of frequency $\omega_L$ of the form $\vec{E}(t)=\vec{E}_0\sin(\omega_Lt)$ is used to define the non hermitian field coupling operator

\begin{equation}
\label{braucheichnicht}
\hat{w}_{\vec{k}} = \frac{ief}{4\pi} \sum_m \sum_{\alpha=1}^3(\vec{a}_\alpha\cdot\vec{E})N_{\vec{k}_\alpha} 
\sum_{\sigma=\pm}\sigma \left| \tilde{v}_{\vec{k}_\alpha^\sigma,m}\right\rangle \left\langle v_{\vec{k},m} \right| .
\end{equation}

In this expression, 

\begin{equation}
\left| \tilde{v}_{\vec{k}_\alpha^\pm,n}\right\rangle= \sum_m (S(\vec{k},\vec{k}_\alpha^\pm))_{mn} \left| {v}_{\vec{k}_\alpha^\pm,m}\right\rangle
\end{equation}

and $S_{mn}$ are matrix elements defined by equation \ref{overlap}.

Replacing the coupling operator $\hat{w}_{\vec{k}}$ with the hermitian form $\hat{w}_{\vec{k}}+\hat{w}^{\dag}_{\vec{k}}$, the equations of motion are obtained:

\begin{equation}
\label{eom}
i\hbar\frac{d}{dt} \left| {v}_{\vec{k},m} \right\rangle = \left( \hat{H}^{0} + \hat{w}_{\vec{k}}(\vec{E})+\hat{w}^{\dag}_{\vec{k}}(\vec{E}) \right) \left| {v}_{\vec{k},m} \right\rangle .
\end{equation}

To account for quasiparticle effects, a scissor operator can be included in equation \ref{eom}, which modifies the eigenvalues of the Hamiltoniana $\hat{H}^{0}$ without modifing the corresponding eigenvectors. 
The equations of motion are then solved to obtain the lattice-periodic part $v_{\vec{k},n}$ of the Bloch states $\left| n\vec{k}\right\rangle$. 
From these, the overlap integrals

\begin{equation}
\label{overlap}
S_{mn}(\vec{k},\vec{k}+\vec{q}_\alpha)=\left\langle v_{\vec{k},m}\middle| v_{\vec{k}+\vec{q}_\alpha ,n}\right\rangle
\end{equation}

are calculated, which are the elements of the matrix $S$, from which the time dependent polarization is calculated as

\begin{equation}
\label{eq:polarisation_num} 
\vec{P}_\alpha =-\frac{ef}{2\pi v}\frac{\vec{a}_\alpha}{N_{\vec{k}_\alpha^\perp}} \sum_{\vec{k}_\alpha^\perp}\Im\left [\sum_{i=1}^{N_{\vec{k}_\alpha}-1}\mathrm{Tr}\ln S(\vec{k}_i,\vec{k}_i+\vec{q}_\alpha)\right ] .
\end{equation}

In a final step, the nonlinear susceptibilities are obtained by postprocessing the polarization in a signal analysis procedure as described in Ref.~\cite{Attaccalite2013}.

\subsection{Computational parameters and electronic groundstate}

The calculation of the nonlinear optical response in frequency-domain from the momentum matrix elements is performed within the density functional theory (DFT) as implemented in the \emph{Vienna Ab initio Simulation Package} (VASP, Version $5.4.4$ \cite{Kresse1996,Kresse1996a}).
The exchange-correlation potential in the formulation of Perdew, Burke, and Ernzerhof \cite{Perdew1996a} and projector-augmented waves  potentials \cite{Bloechl1994} are employed, that include the $2\mathrm{s}^1$, valence electrons in case of lithium, $2\mathrm{s}^2 \, 2\mathrm{p}^4$ in the case of oxygen,  
and $4\mathrm{s}^2 \, 4\mathrm{p}^6 \, 4\mathrm{d}^4 \, 5\mathrm{s}^1$ for  niobium,
respectively.
The basis set for the expansion of the wavefunctions contains plane waves with kinetic energy up to 400\,{eV}. 
The unit cell for the simulation of $\mathrm{LiNbO}_3$ (and $\mathrm{LiTaO}_3$) is rhombohedral and consists of two formula units. 
In this case, the integration in the  reciprocal space is performed on a $6\times 6\times 6$ Monckhorst-Pack grid \cite{Monkhorst1976}, which reflects the symmetry of the unit cell and 
corresponds to 38 $\vec{k}$ points in the irreducible Brillouin zone. 

\begin{figure}[t]
  \includegraphics[width=\linewidth]{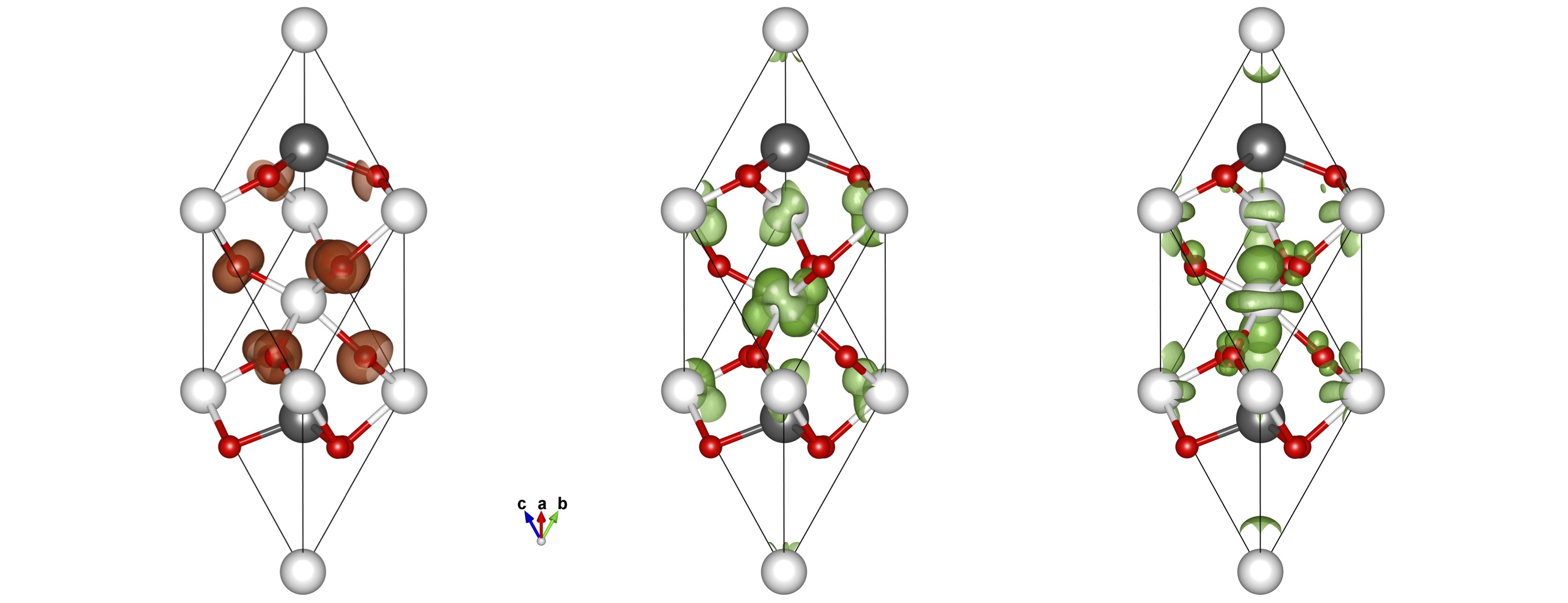}
  \caption{Rhombohedral LiNbO$_3$ unit cell with Li, Nb and O in grey, white and red, 
    respectively. Partial charge densities of the electronic groundstate are shown as 
    isosurfaces (0.011\,{eV/\AA$^3$}). The topmost occupied valence state (left 
    part) and the lowermost unoccupied conduction states (central and right part) 
    strongly resemble the O $2p$ and Nb $4d$ orbitals, respectively.\label{fig:parchg}}
\end{figure}

With these computational parameters, the atomic structure of lithium niobate closely matches that of earlier calculations \cite{Sanna2010,Riefer13,Riefer2016}
and reproduces the experimental values within 1\,\% \cite{Raeuber78,Volk08}. 
The corresponding structure is shown in Fig.~\ref{fig:parchg} together with partial charge densities associated to the valence and (at $\Gamma$ degenerate) conduction band edges. 

\begin{figure}[t]
  \includegraphics[width=8.44cm]{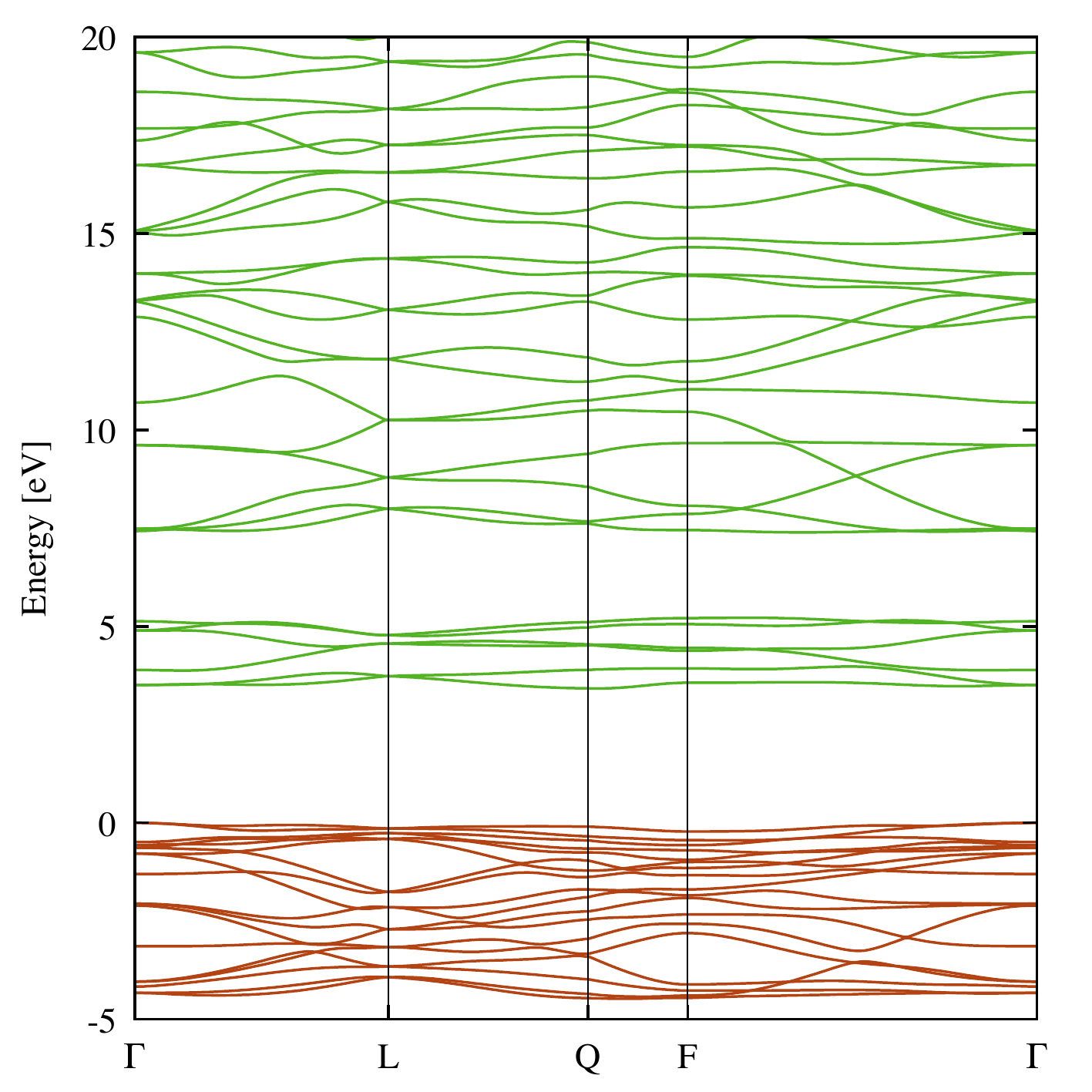}
  \caption{Electronic band structure of LiNbO$_3$. The energies are given relative to the 
    valence band maximum. The direct (indirect) fundamental electronic bandgap amounts to 
    3.52\,{eV} (3.42\,{eV}).\label{fig:bandstructure}}
\end{figure}

The electronic band structure of LiNbO$_3$ is shown in Fig. \ref{fig:bandstructure}. 
It features a very flat dispersion of both valence and conduction states, which is characteristic of the material \cite{Riefer13,Krampf_2021,Mich_Fr}. 
The indirect (direct) fundamental electronic bandgap amounts to 3.52\,{eV} (3.42\,{eV}) which also is in agreement with earlier calculations\cite{Sanna2010,Riefer13}.

\begin{figure}[t]
  \begin{minipage}{0.49\linewidth}
    \includegraphics[width=1.0\linewidth]{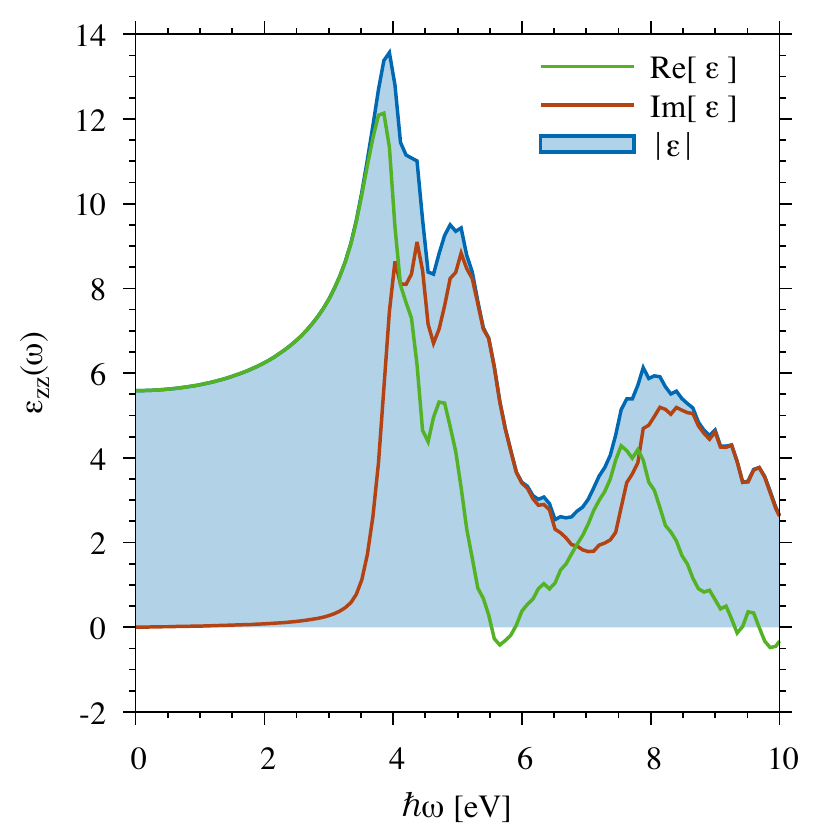}	
  \end{minipage}
  \hfill
  \begin{minipage}{0.49\linewidth}
    \includegraphics[width=1.0\linewidth]{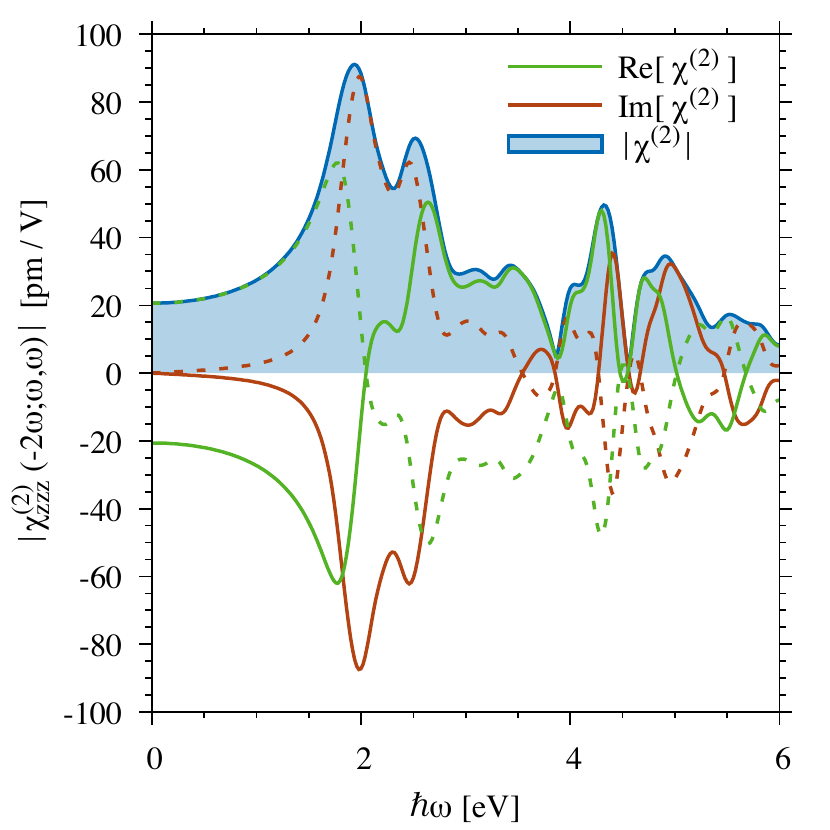}	
  \end{minipage}
  \caption{Left-hand side: $zz$ component of dielectric tensor calculated for ferroelectric 
    LiNbO$_3$ within DFT-PBE in the IPA. Right-hand side: $zzz$ component of the SHG tensor
    of LiNbO$_3$ calculated on the same footing. For better comparison, the real and imaginary 
    part of $\chi^{(2)}$ are also multiplied by $(-1)$ (dashed lines).\label{fig:complexFctn}}
\end{figure}

The rhombohedral unit cell of LiNbO$_3$ contains 64 electrons, which occupy the lowest 32 Kohn-Sham-states. 
In total, 256 bands are employed for the calculation of the linear optical properties because of the slow convergence of the real-part of the dielectric function
with respect to the number of conduction bands. 
The resulting dielectric function (absolute value, along with the real and imaginary part) is shown exemplarily for the $\varepsilon_{zz}$ component in Fig.~\ref{fig:complexFctn}
(lhs). 
The structured peak of the imaginary part at about 4\,eV  represents the most important spectral feature in agreement with previous results\cite{Riefer13}.
The onset of the optical absorption thus corresponds to the bandgap energy.

The calculation of the nonlinear optical response in the time-domain from the time-evolution of the dynamical polarization is performed on the basis of the electronic groundstate 
calculated within the DFT as implemented in the QUANTUM ESPRESSO code \cite{Giannozzi2009,Giannozzi2017} (Version 6.5) using also in this case the Perdew-Burke-Ernzerhof \cite{Perdew1996a} (PBE) functional. 
The set of SG15 ONCV pseudopotentials \cite{Hamann2013,Schlipf2015} is used to describe the electron-core interaction of the involved atoms. 
The wavefunctions are expanded in a plane wave basis up to a cut-off energy of $E=\hbar \left|\vec{k}\right|^2/(2m)=80\,{Ry}\approx$ 1090\,{eV}.
Integration of the reciprocal space has been performed with a $\Gamma$-centered $10\times 10\times 10$  $\vec{k}$-Mesh, which consists of 172 individual $\vec{k}$ points.

The YAMBO \cite{Sangalli2019} code is employed to calculate the optical properties. 
To this end, the DFT (single particle) wave functions calculated with QUANTUM ESPRESSO are imported to build a Kohn-Sham basis set consisting of 22 topmost valence bands and the 43 lowest conduction bands. 
These values have been carefully chosen on the basis of the convergence tests shown in Appendix 1.
Test calculations of the linear optics show an indirect (direct) fundamental electronic bandgap of 3.45\,{eV} (3.52\,{eV}).
The band gap energy values, the band structure and the dielectric function are in overall very good agreement with the VASP calculations.  
Furthermore, additional occupied or unoccupied states only lead to minor changes in the imaginary part of the dielectric constant. 
Six additional states modify the dielectric function by less than 1\,\%.

The optical susceptibilities are explicitly calculated for $\hbar\omega$ representing 64 energy values between 0.2\,{eV} and 6\,{eV}, as well as for an incident electric field with a field strength of $10^{14}$\,{W/m$^2$} and a damping of 0.2\,{eV}. 
The Crank-Nicholson algorithm is used to calculate $\vec{P}(t)$ for 5364 time steps 0.01\,{fs} apart. 
Taking dephasing of eigenmodes introduced by the sudden switch-on of the electric field into consideration, only the last 20.68\,{fs} of the polarization are eligible for frequency analysis, which are used to extract $\vec{P}(2\omega)$ and $\vec{P}(3\omega)$. 
Next, this result is converted into $\chi^{(2)}(-2\omega; \omega;\omega)$ and $\chi^{(3)}(-3\omega; \omega,\omega,\omega)$ for the outbound radiation corresponding to $2\hbar\omega$ and $3\hbar\omega$, respectively, i.e. the harmonic generation of inbound radiation with energy $\hbar\omega$ of second and third order, respectively.

In order to explore the effect of chemical variations in the cationic sublattice on the nonlinear optical response of LiNbO$_3$, the ferroelectric oxides LiTaO$_3$ and KNbO$_3$ were modeled.
The substitution of the transition metal Nb (group 5, period 5) with the heavier, isovalent Ta  (group 5, period 6) leads to the formation of an isomorph crystal, ferroelectric LiTaO$_3$, which likewise crystallizes into the $R3c$ space group and can be modeled by a rhombohedral unit cell containing two formula units. 
The material as well as its properties and applications are in general very similar to LiNbO$_3$. 
However, the Curie \cite{201552576} and melting (1923\,K vs 1526\,K) \cite{korth} temperatures, as well as the coercive fields (17\,kV/cm vs 40\,kV/mm) \cite{1470247} are lower, and the optical nonlinearities in the visible range less pronounced.
In our simulations, LiNbO$_3$ and LiTaO$_3$ are modeled with identical numerical parameters.
Besides a different value of the fundamental electronic bandgap (3.56\,{eV}), LiTaO$_3$ has a very similar, likewise flat electronic bandstructure to LiNbO$_3$, whereby the valence band maximum and the conduction band minimum have O $2p$ and Ta $5d$ character, respectively. 
The dielectric function of LiTaO$_3$ strongly resembles that of LiNbO$_3$ (see Fig.~\ref{fig:complexFctn}, rhs) although all spectral features are slightly shifted to higher energies due to the larger bandgap energy.

The substitution of the alkali metal Li (group 1, period 2) with the heavier, isovalent K (group 1, period 4) leads to a fundamentally different crystal, KNbO$_3$. 
KNbO$_3$ crystallizes, with decreasing temperature, in a cubic, tetragonal, orthorhombic, and monoclinic phase \cite{Schmidt2017}. 
In this work, we focus on the  tetragonal phase (space group $P4mm$, $c/a$ ratio of $1.023$), as it is computationally convenient and because recent studies have shown that it features the largest SHG coefficients in comparison to the other phases \cite{Schmidt2019}. 
The DFT-PBE calculated atomic structure reproduces within1\,\% the experimentally determined structural parameters \cite{ABRAHAMS1973521,201552576}.

\begin{figure}[t]
  \centering
  \includegraphics[width=0.5\linewidth]{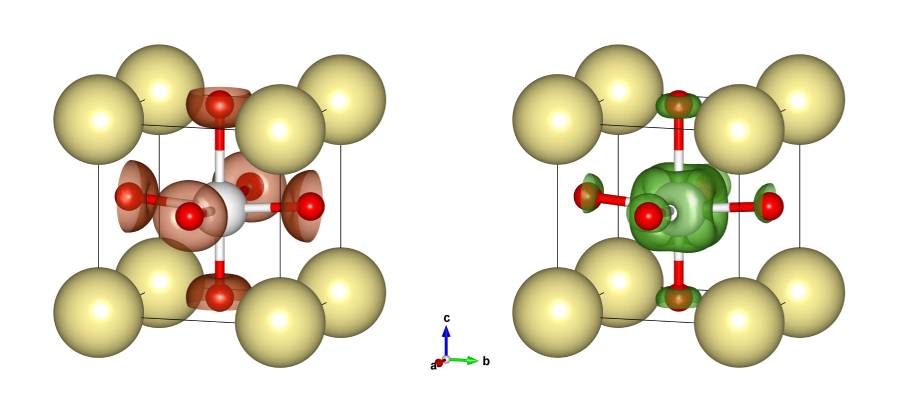}
  \caption{Tetragonal structure of KNbO$_3$, with K, Nb and O atoms depicted in yellow, 
    white and red, respectively. Partial charge densities corresponding to the highest 
    valence band (left-hand side) and the lowest conduction band (right-hand side) are shown. 
    Isosurfaces correspond to a charge density of 0.017\,{eV/\AA$^3$}.}
  \label{fig:knparchg}
\end{figure}

Similarly to LiNbO$_3$ and LiTaO$_3$, the valence band top originates from the anionic sublattice and has O character, while the cations (Nb) determine the conduction band bottom, as shown in Fig.~\ref{fig:knparchg}.

For the calculation of the optical response, a very dense 16$\times$16$\times$16 $k$-point mesh (corresponding to a total of 720\,$k$-points) is necessary. 
For the time evolution of the wavefunctions of the 40 electrons contained in the unit cell, a Kohn-Sham basis made of 15 occupied states and 27 unoccupied states is considered. 
The nonlinear susceptibilities are calculated sampling the energy axis with points at a distance of 0.08\,eV.

\subsection{Experimental setup}
All investigated samples are commercially available X-cut wafers (Surface Net GmbH, Germany).
We use a 50\,fs pulse duration 5\,kHz repetition rate Ti:Sa amplifier to drive an optical parametric amplifier including mixing stages to generate sub-100\,fs pulses from 800\,nm to 1580\,nm. 
The beam diameter is reduced by two concave mirrors and split into two beams using a symmetric Michelson interferometer. 
SHG is measured by placing the crystal in the unfocused beam under  normal incidence. 
Lock-in technique on Si photodiodes are used for detection. 
The fundamental  laser is suppressed by the appropriate dielectric short-pass filters. 
The data are corrected against  a simultaneously acquired reference on z-cut quartz for which the literatrue absolute value of $d_{11}$ = 0.3\,pm/V at 1064\,nm is assuemd \cite{Levine72}. 
We use Miller's constant-delta condition \cite{Miller64} with a Miller delta of $\delta_{11}$ = 1.328 $\times$ 10$^2$\,m/C to account for the dispersion of the second-order nonlinear coefficient of the quartz reference, as was already performed, for example, in Ref.~\cite{Schmidt2019}.

\section{Results and discussion} \label{sec:results}

We start our discussion with the comparison of the SHG spectra obtained with the momentum matrix approach and with the time-evolution of the dynamical polarization. 
As the focus is partially set on the comparison of the two methods, results within the IPA are presented. 
Many-body effects are considered in a second step, based on the quasiparticle shifts reported in Refs.~\cite{Riefer13,Schmidt2019}.
 
After expansion of the wavefunctions from the irreducible part to the full Brillouin zone, and the calculation of their derivatives, the second order optical susceptibility 

\begin{equation}
	\left|\chi^{(2)}\right|^2=\Re{\left[\chi^{(2)}\right ]}^2+\Im{\left [\chi^{(2)}\right ]}^2
\end{equation}

can be estimated in the frequency domain as a sum over momentum matrix elements. 
We find that the two-band contributions are negligible and the three-band contributions play the crucial role.
Fig.~\ref{fig:complexFctn} (lhs) shows the result of this procedure exemplarily for the $\left|\chi^{(2)}_{zzz}(-2\omega,\omega,\omega)\right|$ tensor component. 
The most prominent feature is the first peak at 1.9\,eV with an intensity of 183\,pm/V confirming the strong optical non-linearity of LiNbO$_3$.
In the static limit, $\left|\chi^{(2)}_{zzz}(0)\right|$ is as high as 42\,pm/V. 
The calculated spectrum is in overall good agreement with the results reported in Ref.~\cite{Riefer13}.

\begin{figure}[t]
  \includegraphics[width=\linewidth]{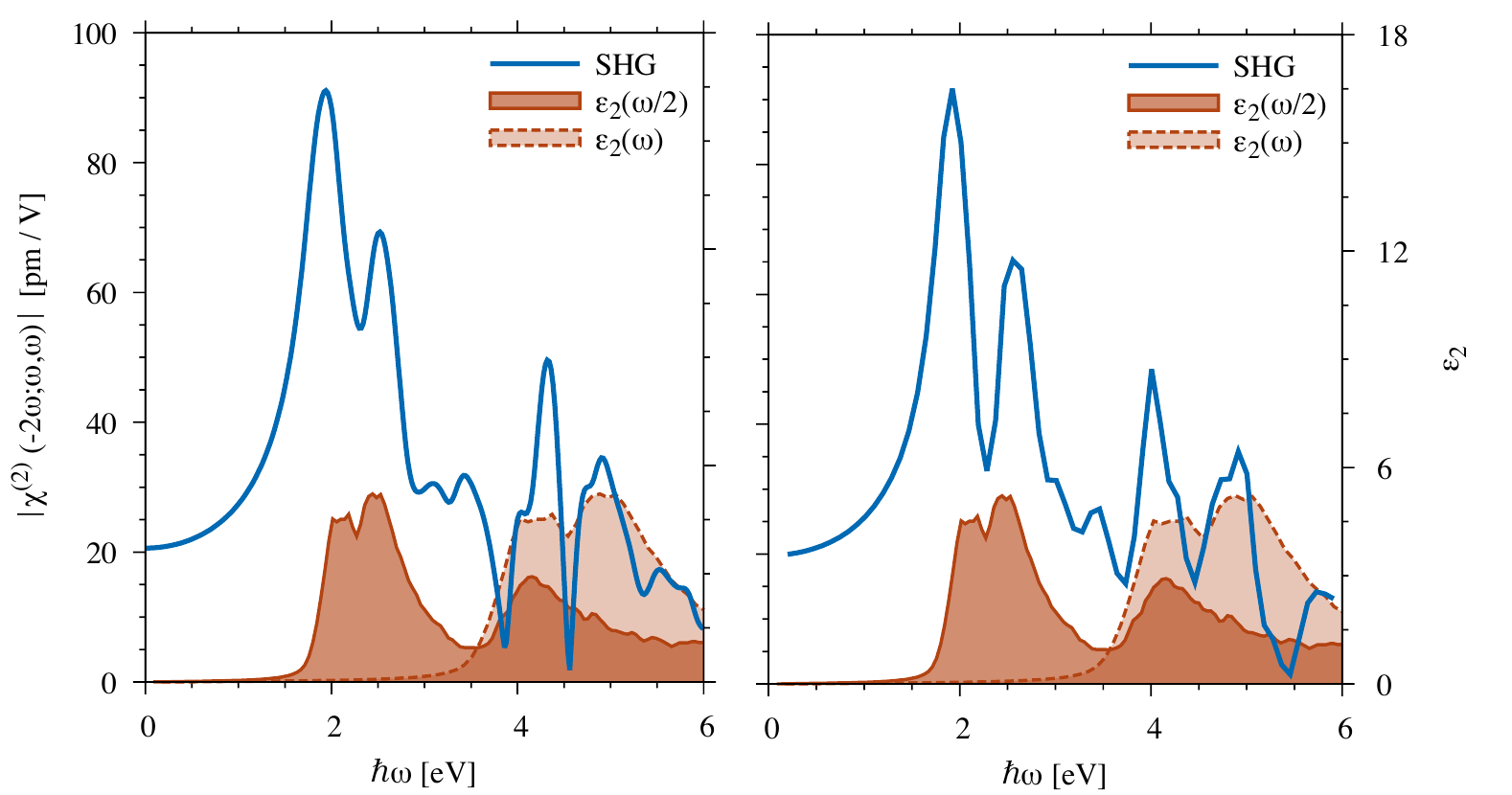}
  \caption{\label{fig:diel_ln} Absolute values of the $zzz$ component of the SHG tensor of 
   LiNbO$_3$ calculated by the momentum matrix elements (left-hand side) and by the 
   time evolution of the polarization (right-hand side). Each of them is compared with 
   the imaginary part of the dielectric function $\varepsilon_{zz}(\omega)$ and its 
   energy-scaled counterpart $\varepsilon_{zz}(\omega/2)$.}
\end{figure}

The position of the resonances in the SHG spectrum can be understood on the basis of the optical absorption. 
To this end, we show in Fig.~\ref{fig:diel_ln} (lhs) the absolute values of the SHG coefficients $\chi^{(2)}_{zzz}$ and the imaginary part of the dielectric tensor $\varepsilon_{zz}(\omega)$ as well as $\varepsilon_{zz}(\omega/2)$. 
The latter is characterized by spectral signatures roughly at energies for which also the SHG spectrum shows the main peaks. 
This means that the SHG spectral features might be understood as two-photon processes. 

In a second step, the SHG spectrum of LiNbO$_3$ is calculated from the time evolution of  the polarization. 
After calculation of $P(t)$ and expansion in a power row up to the 6$^{\mathrm th}$ order, a Fourier analysis yields the spectrum shown in 
Fig.~\ref{fig:diel_ln} (rhs). 
For energies within the fundamental bandgap, two peaks at 1.9\,eV and 2.5\,eV represent the most prominent signatures. 
Both peaks match the spectral features calculated in the frequency domain, both in peak position and intensity.

\begin{figure}[t]
  \centering
  \includegraphics{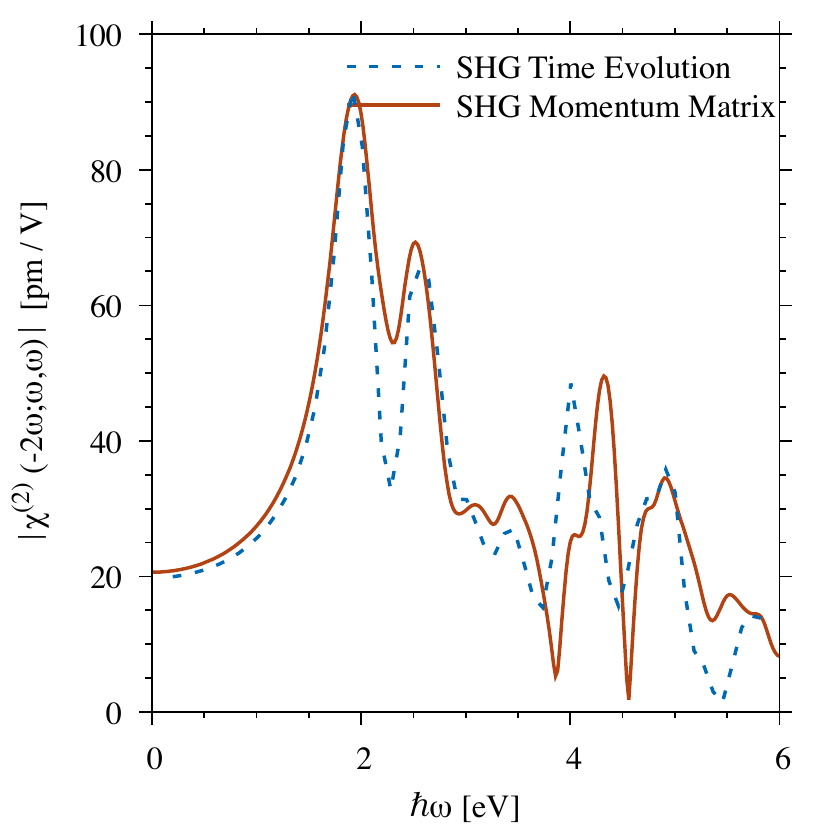}
  \caption{Absolute values of the SHG susceptibility of LiNbO$_3$ (the $zzz$ component is 
    representatively shown) as calculated in the time-domain (dashed blue line) and in 
    the frequency domain (solid red line).}
  \label{fig:shgvaspyambo}
\end{figure}

The direct comparison of the LiNbO$_3$ SHG spectra calculated with the two approaches is shown in Fig.~\ref{fig:shgvaspyambo}. 
In the investigated frequency range, the agreement is excellent concerning both the position and intensity of the spectral features. 
The main difference between the two spectra is represented by the magnitude of the deeps as calculated in the time-domain, which might be due, however, to a different choice of the damping parameter which determines the resonance widths and to the use of different pseudo-potentials. 
The agreement between the results obtained with two fundamentally different approaches is by no means a matter of course, and represents a major mutual validation of the two methods and their implementations.

\begin{figure}[t]
  \includegraphics{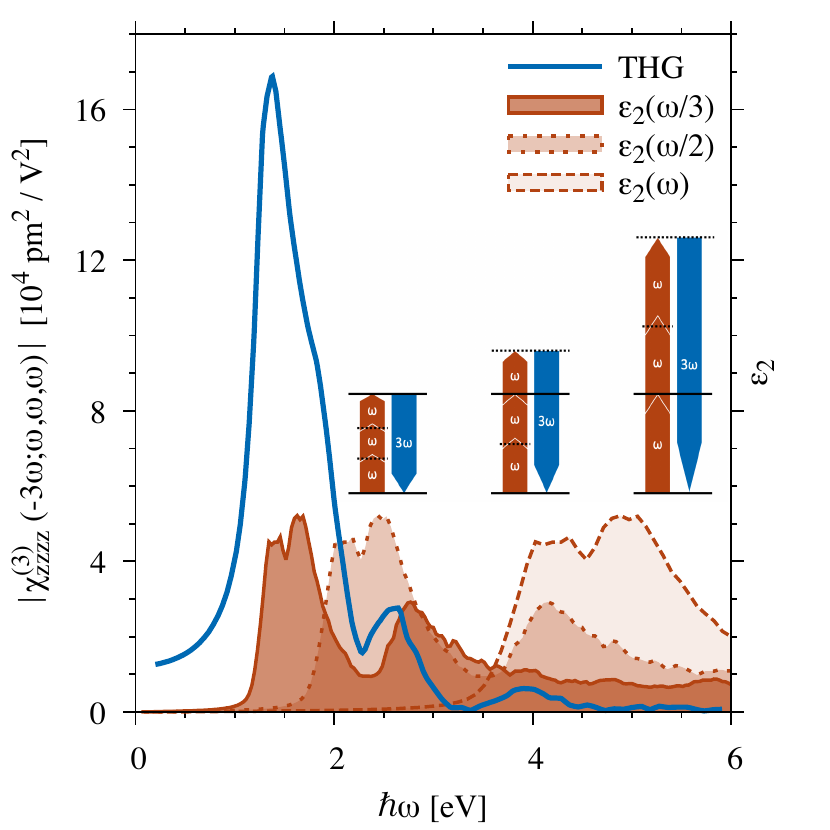}
  \caption{\label{fig:blabla} Absolute values of the $zzzz$ component of the THG tensor of 
    LiNbO$_3$ as calculated from the time evolution of the polarization. The imaginary part 
    of the corresponding component of the dielectric tensor $\epsilon_{zz}(\omega)$ and 
    its energy-scaled counterparts $\epsilon_{zz}(\omega/2)$ and $\epsilon_{zz}(\omega/3)$ 
    are reported for comparison.}
\end{figure}

The major advantage of the time-domain approach over the frequency-domain approach is the simultaneous calculation of the higher harmonics. 
Fig.~\ref{fig:blabla} shows exemplarily the $zzzz$ component of the THG tensor of LiNbO$_3$. 
It is characterized by a dominant peak at 1.4\,eV with a very strong intensity of about 1.7$\cdot$10$^5$\,pm$^2$/V$^2$, a minor feature at 2.6\,eV and a small shoulder at 3.9\,eV. 
A comparison with the corresponding component of the dielectric function shows clearly that the first peak is due to three-photon processes, the second feature to two-photon processes, and the minor shoulder with roughly the bandgap energy is due to one-photon resonances.
These are schematically shown in the figure inset.

Next, we compare the calculated optical response of LiNbO$_3$ with that of LiTaO$_3$ and KNbO$_3$ to determine how the substitutions of the transition metal and of the alkali metal affect
the optical properties.

\begin{figure}
		\includegraphics[width=1.0\linewidth]{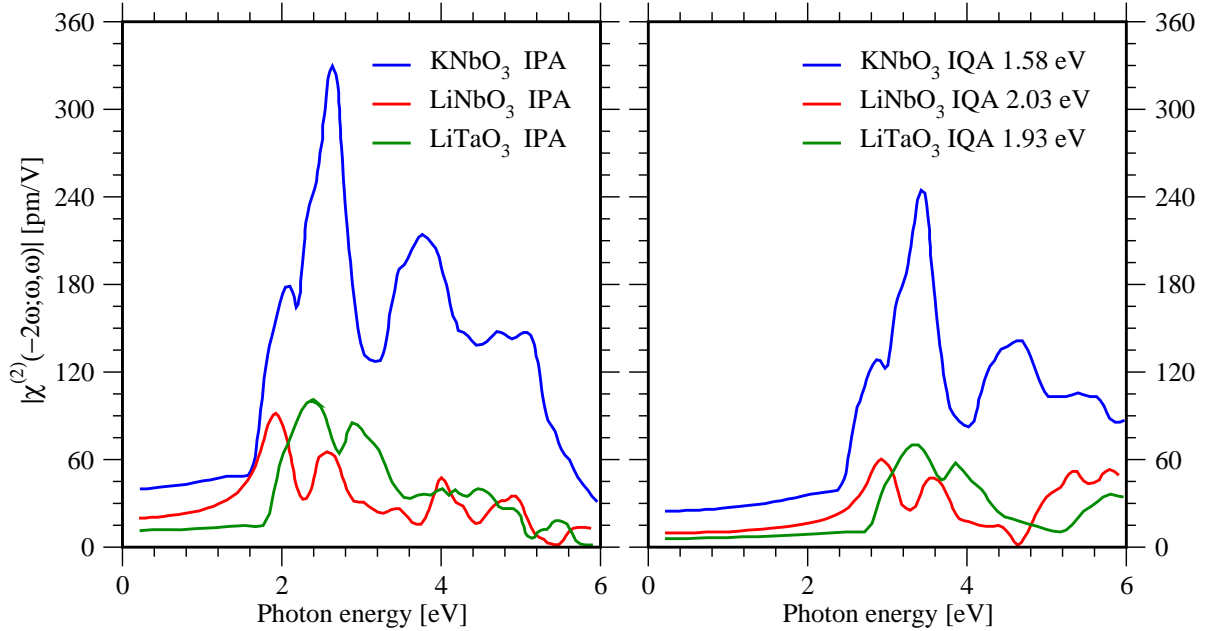}	
	\caption{Absolute values of the $zzz$ component of the SHG tensor in the IPA (left-hand side) 
          and in the independent quasiparticle approximation (right-hand side) calculated for different
          ferroelectric oxides.}
	\label{fig:shg_FE}
\end{figure}

\begin{figure}
		\includegraphics[width=0.8\linewidth]{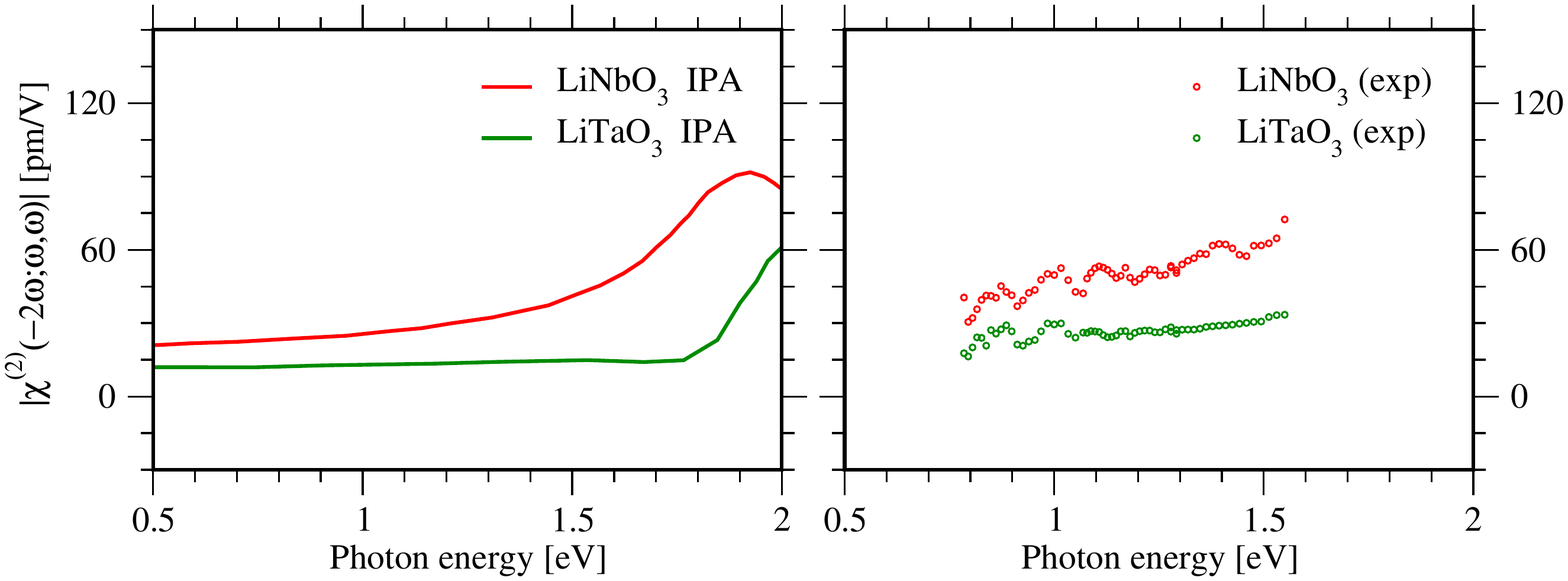}	
	\caption{Absolute values of the $zzz$ component of the SHG tensor of LiNbO$_3$ and LiTaO$_3$ 
          as calculated by DFT in the IPA (left-hand side) and measured (right-hand side).}
	\label{fig:experiment}
\end{figure}

Figure~\ref{fig:shg_FE} (lhs) shows a comparison of the $\chi^{(2)}_{zzz}$ component of the second-order optical susceptibility calculated for the investigated crystals, while Fig.~\ref{fig:experiment} shows a comparison of calculated and measured SHG spectra for LiNbO$_3$ and LiTaO$_3$. Notably, the calculations correctly predict the spectral depence of the measured values in the experimentally accessible range from 0.78\,eV to 1.55\,eV and also the magnitude agrees exeptionally well, i.e., within a factor of two.

The SHG coefficients of LiNbO$_3$ and LiTaO$_3$ are similar concerning both peak position and intensity, although the LiTaO$_3$ spectrum is blueshifted due to the higher bandgap. 
We observe, however, that the energy shift between the two spectra (the distance between the first peak of each spectrum is of about 0.45\,eV) is higher than the difference between the two bandgap energies (about 0.15\,eV). 
This suggests that not only the band edges but also additional states close to the gap contribute to this spectral feature.
As the bandstructure of the two ferroelectrics are very similar, the energy shift might be explained by the different lattice  parameters of the two compounds. 
In particular, the two spectra differ in the low energy region before the first peak. 
While LiNbO$_3$ is characterized by a continuous increase, the SHG coefficient of LiTaO$_3$ remains below 15\,pm/V till 1.8\,eV, followed by a steep gradient which brings the optical susceptibility to 101\,pm/V at 2.4\,eV. 
This behavior, which also holds true when quasiparticle effects are included in the calculation, might explain why lower SHG coefficients are measured for LT in the visible range.

The SHG spectrum of KNbO$_3$ is fundamentally different, from those of LiNbO$_3$ or LiTaO$_3$. 
In the long-wavelength limit, the $zzz$ component of the SHG coefficient (ca. 40\,pm/V) is much larger than in the case of LiNbO$_3$ or LiTaO$_3$. 
After a steep gradient beginning at 1.5\,eV, the SHG coefficient reaches a global maximum of 330\,pm/V at 2.6\,eV. 
Thus, the $zzz$ component of the SHG coefficient as calculated within the time-domain is in qualitative agreement with the calculations presented in Ref.~\cite{Schmidt2019} and based on the momentum-matrix approach.

Figure~\ref{fig:shg_FE} (rhs) shows how the $\chi^{(2)}_{zzz}$ component of the second-order optical susceptibility of the investigated crystals is affected by quasiparticle effects. 
Besides an expected blueshift of half the value of the shifts (indicated in the figure), the quasiparticle effects lead to an overall reduction of the signal intensity.
A similar effect has been predicted by Riefer \textit{et al.}~\cite{Riefer13} for LiNbO$_3$. 
Nevertheless, no major redistribution of the spectral weights is observed, and the relative differences between the spectra closely mirror those described for the IPA spectra.

\begin{figure}[t]
	\begin{minipage}{0.49\linewidth}
		\includegraphics[width=1.0\linewidth]{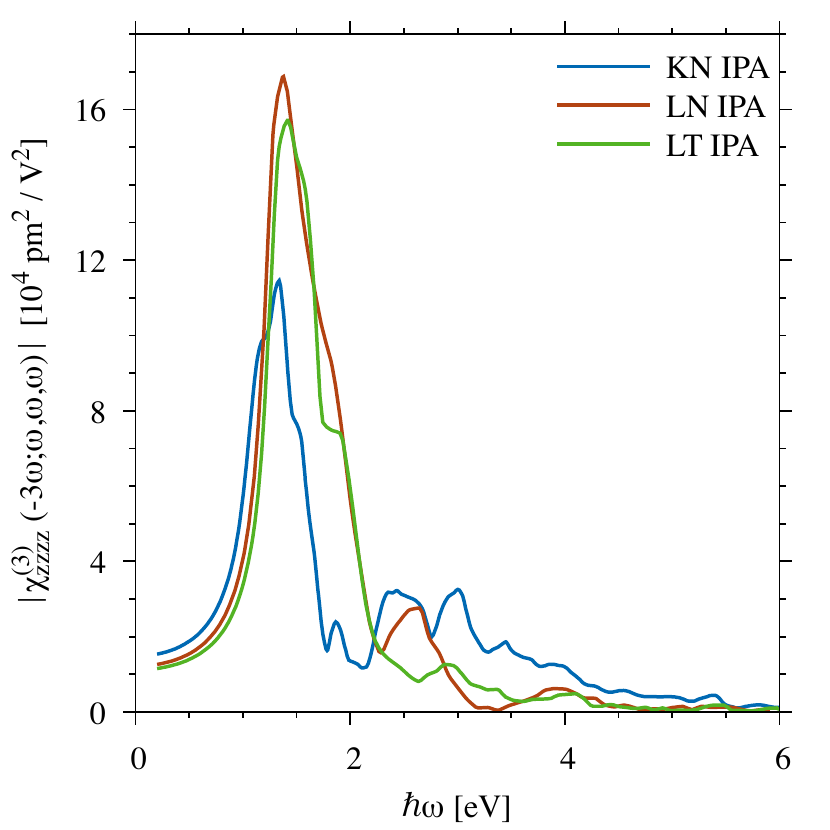}	
	\end{minipage}
	\hfill
	\begin{minipage}{0.49\linewidth}
		\includegraphics[width=1.0\linewidth]{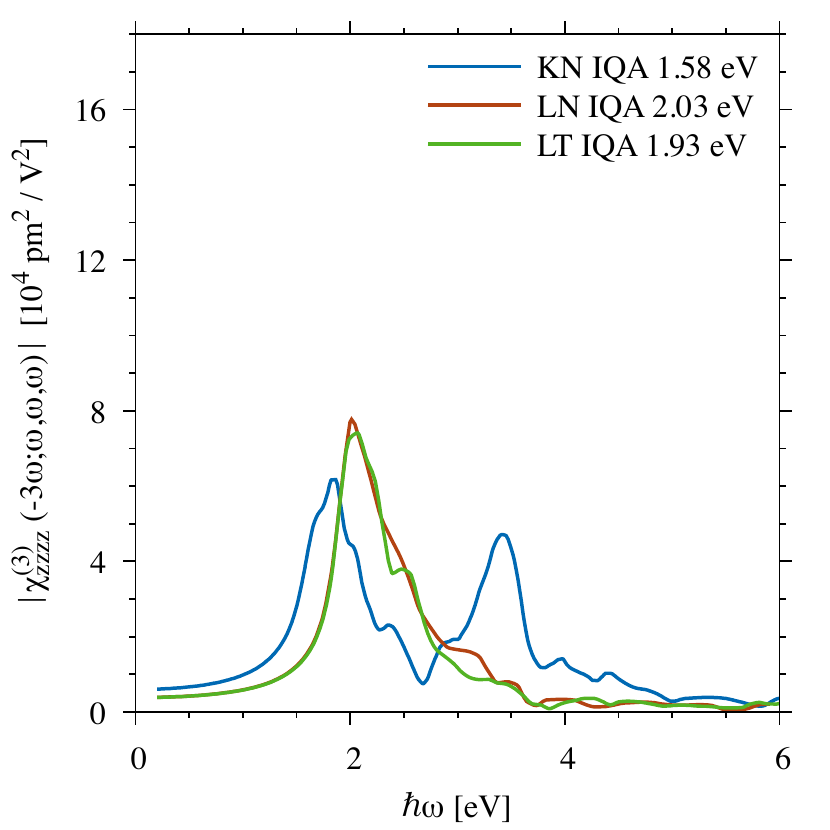}	
	\end{minipage}
	\caption{Absolute values of the $zzzz$ component of the THG tensor in the IPA (left-hand side) 
          and in the independent quasi-particle approximation (right-hand side) of different
          ferroelectric oxides.}
	\label{fig:thg_FE}
\end{figure}

The $\chi^{(3)}_{zzzz}$ component of the THG spectra (reported in Fig.~\ref{fig:thg_FE}, lhs) as calculated in the IPA for the investigated oxides has a rather similar behavior for all
crystals and does not feature the differences observed for the corresponding SHG tensor component. 
All spectra are dominated by main THG peak between 1\,eV and 2\,eV, whose onset mirrors the order of the bandgap energies of KNbO$_3$, LiNbO$_3$ and LiTaO$_3$. 
Yet, the third-order energy differences amount to one third of the bandgap energy, so that the three main peaks are rather close to each other. 
The height of the maxima grows from KNbO$_3$ (115000\,pm$^2$/V$^2$)  to LiNbO$_3$ (169000\,pm$^2$/V$^2$) which is the reverse order of the SHG maxima. 
On contrary, the order of the THG activity in the long-wavelength limit mirrors that of the corresponding SHG component, with the highest value for KNbO$_3$ (15500 pm$^2$/V$^2$) and the lowest for LiTaO$_3$ (11500\,pm$^2$/V$^2$).

When quasiparticle effects are considered (Fig.~\ref{fig:thg_FE}, rhs) the intensity  of the THG signatures is drastically reduced by about 50\% of the IPA value.
Yet, LiNbO$_3$ and LiTaO$_3$ do not undergo large redistributions of the relative spectral weights.
However, due to the quasiparticle-induced shifts the THG spectra of LiNbO$_3$ and LiTaO$_3$ become almost indistinguishable in the visible range. 

Differently than in the case of LiNbO$_3$ and LiTaO$_3$, the quasiparticle effects strongly modify the KNbO$_3$ THG spectrum.
The intensity of the main peak is somewhat less affected than in the case of the other ferroelectrics, and, more important, the intensity of the peak at 3.4\,eV is strongly increased to 47000\,pm$^2$/V$^2$.
This is an unusual feature, as the general decrease of the optical susceptibility upon widening of the fundamental gap is a known effect based on the sum rules for the harmonic susceptibilities \cite{Saarinen2002,Panday2017}. 

We remark that neither the crystal local fields nor the electron-hole interaction have been considered in our calculations. 
While local fields are expected to reduce the optical nonlinearities, excitonic effects typically enhance them \cite{Luppi2010,Luppi2016}, so that some error cancelation might be expected.  
Yet exciton effects are in some cases not strong enough enought to compensate the quasi-particle effects in THG \cite{PhysRevB.95.125403}.


\section{Conclusions}
The nonlinear optical response of the ferroelectric oxides LiTaO$_3$, LiNbO$_3$, and KNbO$_3$ is investigated from first principles and compared with specifically performed measurements of 
the SHG spectra of LiTaO$_3$ and LiNbO$_3$. 
To this end, two different approaches, based on the calculation of the momentum-matrix elements and of the dynamical polarization in its Berry phase formulation, are employed, respectively. 
While the second order optical susceptibility is calculated in the frequency domain, second-order and third-order nonlinearities are calculated within the time-domain. 
The excellent agreement between the two approaches concerning the calculated second order optical susceptibility validates both methods against each other, and allows to reproduce the experimental results as well as earlier results where available.
KNbO$_3$ shows the highest SHG coefficients, while all three materials are comparable concerning THG. 
The main spectral features of all investigated ferroelectrics can be explained by parametric processes and multi-photon resonances. 
The consideration of quasiparticle effects blueshifts the spectra  without affecting the relative differences between them, and reduces the optical nonlinearities. 
This reduction is particularly severe concerning  THG, which is reduced by about 50\% with respect to the IPA calculations. 
The only exception is an anomalous peak at 3.4\,eV in the THG of KNbO$_3$, which gains in intensity upon consideration of quasiparticle effects. 
Our calculations furthermore reveal that LiNbO$_3$ and LiTaO$_3$ have distinct SHG spectra but almost indistinguishable THG spectra.
It must be therefore possible by means of LiNb$_x$Ta$_{1-x}$O$_3$ solid solutions to tune the  crystal's second order optical susceptibility without affecting the third harmonic generation.

Due to the efficiency of the time-domain approach and the possibility to calculate higher harmonics, this method seems particularly suitable to investigate complex nonlinear processes from first principles. 
This might be particularly important, e.g., to understand the mechanisms underlying the generation of a high directional light supercontinuum upon infrared irradiation recently observed in organotetrel molecular crystals \cite{Rosemann2018}.

\section*{acknowledgement}
This work is supported by the German Science Foundation (DFG) through the research groups FOR2824 and FOR5044. 
Calculations for this research were conducted on the Lichtenberg high performance computer of the TU Darmstadt and at the H{\"o}chstleistungrechenzentrum Stuttgart (HLRS). 
The authors furthermore acknowledge the computational resources provided by the HPC Core Facility and the HRZ of the Justus-Liebig-Universit{\"a}t Gie{\ss}en.

\section{Appendix: Convergence of numerical parameters}

\begin{figure}
	\includegraphics[width=\linewidth]{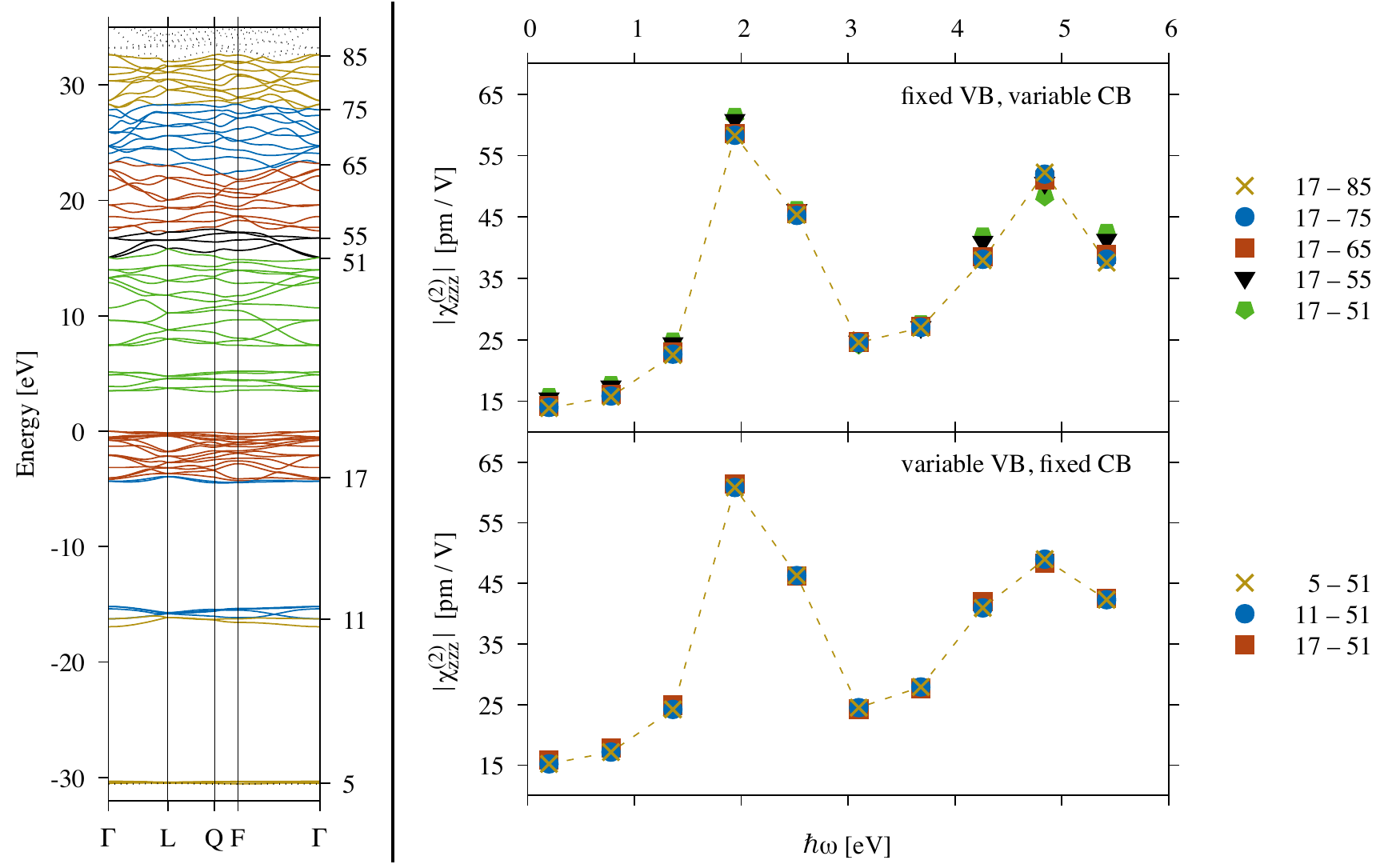}
	\caption{Lhs: Band structure of LiNbO$_3$, in which the energy relative to the 
          valence band maximum is given on the left axis and the corresponding band indices 
          (cf. rhs) are given on the right axis. Rhs: Resulting SHG spectra of LiNbO$_3$ 
          calculated with different Kohn-Sham basis sets. In the bottom (top) half of the 
          figure, the lower (higher) boundary of the band range of the Kohn-Sham basis is 
          varied, while the higher (lower) boundary is fixed. In each case, the 
          corresponding band indices are given in the key. These calculations are all 
          performed with a $4\times 4\times 4$ grid of $k$-points. VB and CB stay for 
        valence and conduction bands, respectively.}
	\label{fig:conv_bands}
\end{figure}

\begin{figure}
	\includegraphics{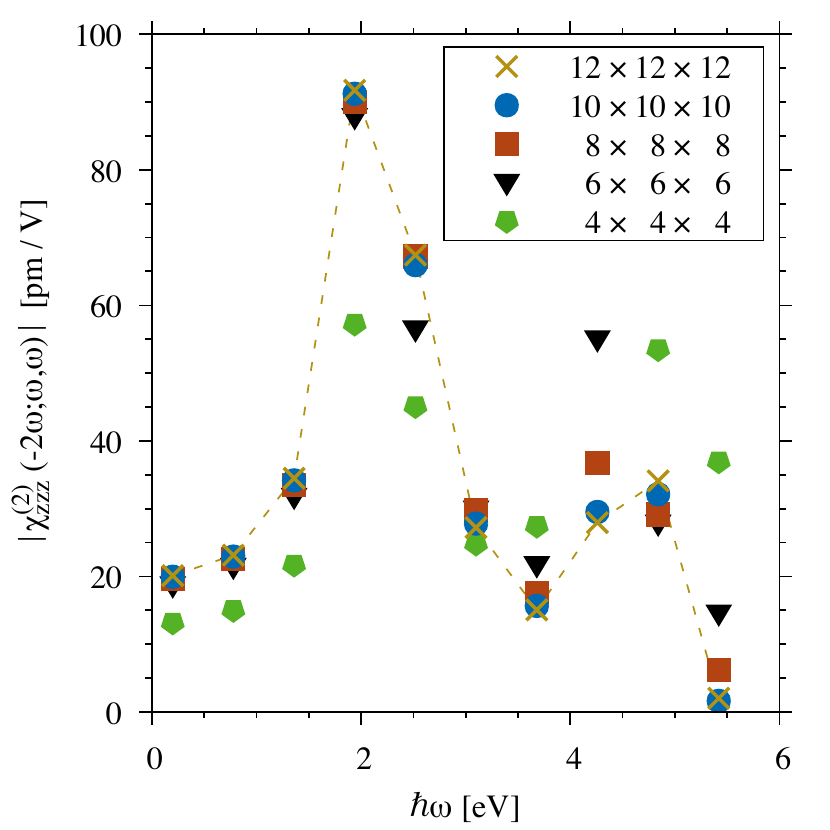}
	\caption{Resulting SHG spectra of LiNbO$_3$ calculated with different number of 
          subdivisions regarding the discretization of the Brillouin zone. All calculations 
          are performed with a Kohn-Sham basis consisting of bands $11-75$.}
	\label{fig:conv_kgrid}
\end{figure}

\bibliography{citations}

\end{document}